\documentclass[a4paper,12pt]{article}
\usepackage[authoryear]{natbib}
\usepackage[T1]{fontenc}

\usepackage{graphicx}
\usepackage[obeyspaces]{url}
\usepackage{fourier} 
\usepackage{upquote}

\usepackage{amsmath}
\usepackage{comment}
\usepackage{listings}
\author{Frank Borg \thanks{Kokkola University Consortium Chydenius, Finland. Email: \url{borgbros@netti.fi}}}
\title{Analyzing biosignals using the R freeware (open source) tool}

\begin{document}

\maketitle

\begin{abstract}

For researchers in electromyography (EMG), and similar biosginals, signal processing is naturally an essential topic. There are a number of excellent tools available. To these one may add the freely available open source statistical software package R, which is in fact also a programming language. It is becoming one of the standard tools for scientists to visualize and process data. A large number of additional packages are continually contributed by an active community. The purpose of this paper is to alert biomechanics researchers to the usefulness of this versatile tool. We discuss a set of basic signal processing methods and their realizations with R which are provided in the supplementary material. The data used in the examples are EMG and force plate data acquired during a quiet standing test. 

\end{abstract}

\section{Introduction}
\label{SEC:Intro}

In biomechanics the application of electromyography is especially a field rife with various signal processing methods. One reason for this is the complexity of the EMG signal which requires processing if one wants to proceed beyond the simple level of on-off interpretation. A number of standard processing tools can be found in EMG softwares by commercial vendors. Researchers are usually also interested in new, or modified processing methods, which may take long to implement in commercial softwares if it happens at all. In this case one can choose ''prototyping'' tools like \textsc{LabView, Mathcad, Matlab, Octave, SciLab}, and so on. Over the the years we have for instance frequently used \textsc{Labwindows CVI} (a C-based graphical programming tool) and \textsc{Mathcad} for data acquisition, visualization and analysis. One common problem in data intensive projects is that the data and analysis documents get scattered among computers and data disks. Our solution is to transfer all data to a single database (in our case based on \textsc{MySql}). Subsets of data can then be selected by \texttt{sql} searches and exported as \texttt{csv}-tables or similar. The salient point is that all the analyses can be gathered in one single script file using the R-tool \citep{R2013}. Running this script generates all the graphs, figures and statistical reports that one may need for the project. This script is easily modified when needed and can be shared among collaborators and other researchers. We believe that R provides a useful platform for defining and using various signal analyzing algorithms in EMG studies, which may stimulate further refinements and developments through collaborative efforts. As noted by \citep[Preface]{Everitt2010}: 

\begin{quote}
(...) R has started to become the main computing engine for statistical research (...) For reproducible piece of research, the original observations, all data processing steps, the statistical analysis as well as the scientific report form a unit and all need to be available for inspection, reproduction and modification by the readers.
\end{quote} 

With this paper and the supplementary material we hope to give a demonstration of the potential uses of R in biosignal processing.


\section{What is R?}

R is an interpreted functional programming language for statistical computing and data visualization created by Ross Ihaka and Robert Gentleman (University of Auckland, New Zealand). R is part of the GNU project and since 1997 it is developed by the \emph{R Development Core Team}. The version R 1.0.0 was released on 29 February 2000. The official project website is \url{www.r-project.org} where one can naturally download the software (binaries, or source code) and manuals \citep{Venables2009}. R is an implementation of the S language developed at the AT\&T Bell Laboratories around 1975 (by Rick Becker, John Chambers and Allan Wilks), and is influenced by the \textsc{Lisp} dialect \textsc{Scheme} created at the MIT AI Lab, also around 1975 (by Guy L Steele and Gerald Jay Sussman). A large part of the software is written in the R language itself, but core functions are written in C and \textsc{Fortran}. As an interpreted language R may be slower than say pure C-programs. Klemens, who advocates C as general scientific programming tool, gives an example with running the Fisher test 5 million times: the speed relation came out as about 1:30 in favour of C \citep{Klemens2009}. Thus in speed-critical cases, such as where we have large simulation samples, it may be necessary to revert to compiled C programs. However, in many other cases the versatility of R will outweigh the possible gains from optimizations for speed. The R tool is continually extended by packages contributed by an active community; there are presently (2014, July) about 5700 packages available, see \url{http://cran.r-project.org/web/packages/}.

Downloading and installing the R software on your computer should require nothing more than basic computer skills. A useful companion is the R scripting tool \textsc{NppToR} (\url{http://sourceforge.net/projects/npptor/}) which is an adds-on to the editing tool \textsc{Notepad++} which can be downloaded from \url{http://notepad-plus.sourceforge.net/uk/site.htm}. Popular graphical editors for R are, among others, \textsc{Tinn-R} and \textsc{RStudio} available from \url{http://nbcgib.uesc.br/lec/software/editores/tinn-r/en} and \url{http://www.rstudio.com}.  

Since documentation about R is easily found on the web we will move on to describe what we can do with the software.

\section{Importing data}

The first thing we want to do is to get our data into the workspace. R can access various databases directly but the most typical situation is where we have, say, EMG data exported on a text format. We use R as console program, so the command for reading a simple ASCI table of data from a file into variable \verb|M| using the \verb|read.table()| function is:

\begin{verbatim}

   M <- read.table("C:/.../myData.asc", header = FALSE)
   
\end{verbatim}

where the first argument contains the path to your data file. If the columns are separated for instance by \verb|";"| (instead of space) we have to add to the arguments \verb|sep = ";"|. Note also that the general assignment operator in R is an arrow \verb|<-| and not \verb|=| (which has a more restricted use in R). We have assumed that the data table had no header. In this case column names can added by the command,

\begin{verbatim}

   names(M) <- c("colName1", ..., "colNameN")
   
\end{verbatim}   

if there are $N$ columns. You can then refer, for instance, to the first column as \verb|M$colName1|. The other way to extract the first column is as \verb|M[,1]|. One can inspect the content of the table read into \verb|M| by the command \verb|edit(M)|. Finally you can quit R by the command \verb|q()|. Instead of typing and running the commands at the console in a sequence, you can collect them in a script file (an ordinary text file) with the extension R, named say \verb|myScript.R|, and run it by the command \verb|source("C:/.../myScript.R")| where the argument contains the path to your script file.

\section{Analysing data}
\subsection{Plotting data}
The next thing you want to do is to inspect your data. For a quick plot of the data in the example above use the command, 

\begin{verbatim}
plot(M$colName1, type = "l") 
lines(M$colName2, type = "l")
\end{verbatim}
  
The second line adds the graph of the second column to the same plot. The \verb|plot| function has various arguments for controlling colours, titles, scales, and so on. Information about the function \verb|plot| is obtained by the command \verb|?plot|. An interesting feature is that with a few lines of code it is for example possible to plot histograms/graphs of hundreds of variables and print them to a pdf document for quick browsing on the computer.

After these preliminaries we can move on to the processing of data. In this connection we must explain how we define new functions in R.

\subsection{Basic signal processing -- some examples}

\subsubsection{User defined functions}

In most programming tasks it us convenient to able to define your own functions. We give a simple example how it works in R. Let us say that we want to sum the elements of vectors using a function \verb|sumVec(V)| which returns the sum and mean of a numeric vector \verb|V|. This can be defined in R by:

\begin{lstlisting}
sumVec <- function(V, start = 1){
	n <- length(V)
	sm <- 0
	for(i in start:n){
		sm <- sm  + V[i]
	}
	mn <- sm/(n - start + 1)
	return(list(sum = sm, mean = mn))
}
\end{lstlisting}

We can call this as \verb|result <- sumVec(V)| which puts the sum and mean of the elements of the vector \verb|V| into to a variable \verb|result|. Here we have also demonstrated the very useful \verb|list| structure in R. The sum (\verb|sm|) and mean (\verb|mn|) are put into a list with respective (arbitrary) names \verb|sum| and \verb|mean|. The variables are then referenced as \verb|result$sum| and \verb|result$mean| after the call 

\begin{lstlisting}
result <- sumVec(V). 
\end{lstlisting}

It is an advantage to have a function return a \verb|list| since it is then easy to modify the function by adding new variables to the output \verb|list| without affecting previous uses of the function. The function example also  illustrates an other aspect of R function; that is, we may have arguments with default values, like \verb|start| as in the example. If the argument is not listed, as in 

\begin{lstlisting}
result <- sumVec(V), 
\end{lstlisting}

it will use the default value (\verb|start = 1|). In many functions we have set as default \verb|plot = FALSE| for a variable \verb|plot|, which means that the results will not plotted unless one adds \verb|plot = TRUE| to the arguments.

Functions can also have other functions as arguments, as for example in the case of \verb|EMG_spec| below. This example also illustrates the similarity with the common C and \textsc{Java} syntax. User defined functions can be collected in a separate file \verb|myFunctions.R| which can be made active (loaded into the workspace) by calling it using 

\begin{verbatim}
source("C:/.../myFunctions.R") 
\end{verbatim}

This corresponds to the \verb|include| statement of header files and source files in C programming. The EMG processing functions to be described below are collected in the file \verb|EMGfuns.R| in the supplementary material to this paper. The hash-symbol \verb|#| is used for comment lines in the R scripts.

\subsubsection{Simulated data}

Is useful to have access to various test data, and as example we have implemented a a standard algorithm \citep[pp.70-71]{Hermens1999} in \verb|EMG_sim| which generates simulated EMG data of desired length. This function returns a \verb|list| where the data is contained in the component \verb|sim|. For instance one can probe the spectrum function using the test data as follows:

\begin{lstlisting}
mysim <- EMG_sim(3000)
myspec <- EMG_spec(mysim$sim, plot = TRUE)
\end{lstlisting}

\subsubsection{Rectification, RMS and turns}

Rectifying raw data means simply taking the absolute value of the elements. For EMG data it may be useful to first subtract any offset (bias) from the raw data. Thus the rectification of \verb|V| could be written as \verb|abs(V - mean(V))| and the \verb|EMG_rect| function in \verb|EMGfuns.R| becomes very simple. It is noteworthy that effect of the rectification of EMG is similar to the rectification of AM radio waves whose purpose is to enhance the low frequency components which encode the voice signals. For EMG the low frequency ''voice'' part corresponds to the encoded force \citep{Borg2007}.

A bit less trivial from the programming point of view is the \verb|EMG_rms| function which computes the Random Mean Square of the EMG and can basically be represented as  

\begin{equation}
\sqrt{ \frac{1}{\Delta T} \int_{t - \Delta T/2}^{t + \Delta T/2} EMG(u)^2 du}. 
\end{equation}

To write these sorts of filtering or enveloping functions it is convenient to use the built-in \verb|filter(V, filtc, ...)| function. This takes input data \verb|V| and outputs a vector with elements

\begin{equation}
	y[i] = \mathrm{filtc}[1] \cdot V[i+o] + \cdots + \mathrm{filtc}[p] \cdot V[i+o-(p-1)],
\end{equation}

where \verb|filtc[i]| are the filter coefficients. The offset $o$ depends on the argument \verb|sides| such that \verb|sides = 2| corresponds to zero lag with $o = (p-1)/2$ if $p$ is odd. For the moving average one uses \verb|method = "convolution"|. The function \verb|EMG_rms| has an argument \verb|DT| which determines the size in milliseconds of the moving window over which one calculates the RMS. This function also illustrates a special feature of R: the use of the \verb|'...'| argument.

\begin{lstlisting}
EMG_rms <- function(V, sampFreq = 1000, DT = 250, plot = FALSE, ...) 
{ #part of the function declaration
	rectV <- sqrt(filter(rectV^2, filter1, sides = 2, 
	method = "convolution", ...))
#rest of the function declaration
}
\end{lstlisting}

In this case it means that we can pass arguments to the \verb|filter| function employed by \verb|EMG_rms|. For instance, \verb|filter| has an argument called \verb|circular|, and via \verb|EMG_rms(..., circular = TRUE)|, we can pass a value (\verb|TRUE| in this case) to this argument of the function \verb|filter|.

A version of the classical method of counting turns \citep{Willison1963,Willison1964} is implemented by the function \verb|EMG_turns| and it also uses a moving window \verb|DT| over which one sums the number of turns. The implementation \verb|EMG_wturns| is maybe closer to the original idea by Willison but the practical difference between the two seems small. The turns functions return a structure 

\begin{verbatim}
list(turns.ps = turns_per_sec, turns.where = turns).
\end{verbatim}

The variable \verb|turns.where| contains the time indexes where the turns are counted. (This is also shown in the plot of the function.) This data may be of interest when, for instance, one wants to calculate an entropy metric for the signal.

\subsubsection{Time-frequency domain}

Part of the inspection of the EMG signal is to study its frequency properties. This is typically performed by calculating the power spectrum of the data. For this purpose one subdivides the original times series into blocks of some time length $\Delta T$, then calculates the power spectrum for these and take their mean as the final power spectrum. For the subdivision one normally use a 50 \% overlap which further suppresses the variance of the final spectrum estimate \citep{Press2002}. This method is implemented in the function \verb|EMG_spec|. It uses a default windowing of the data by the filter \verb|filtWelch|. The window size $\Delta T$ is given by the argument \verb|DT| in milliseconds. The nominal frequency resolution is then given by $\Delta f = 1/ \Delta T $. The function outputs a \verb|list|  with the power density estimate in \verb|psd|, which also contains mean (MNF) and median frequency (MDF) in \verb|meanf| and \verb|medianf|. The time-frequency methods naturally rely on the \emph{Fast Fourier Transformation} (FFT) which in R is called by \verb|fft|. Using the \emph{Short Time Fourier Transformation} (STFT), which applies the FFT to subintervals of the time series, we can, for instance, calculate how the mean frequency varies with time. This is implemented by the function \verb|EMG_stft_f|. One use of mean/median frequency is for the study of muscle fatigue as a function of time, which is often associated with a decrease in mean frequency \citep{Lindstrom1977}. 

Using the \verb|fft| transform we can filter EMG signals by suppressing the higher frequency components. The method employed in \citep{Borg2007} is here implemented by the function \verb|EMG_bw0|. It first calculates the Average Rectified Value, then applies \verb|fft| which gives the Fourier coefficients $c(f_k)$. These are multiplied by a filter factor,

\[
c(f_k) \mapsto {\tilde c}(f_k) = \frac{c(f_k)}{1 + \left(\frac{f_k}{f_c}\right)^n},
\]

where $f_c$ is the low pass cut-off frequency of the filter and $n$ is the order of the filter. Finally we obtain the filtered signal by applying the inverse \verb|fft| to ${\tilde c}(f_k)$. This is basically a zero-lag version of the Butterworth filter. With $n = 4$ and $f_c = 1$ Hz we obtained quite a good correspondence between gastrocnemius EMG and the muscle force as expressed by the anterior-posterior COP (center of pressure) during quiet standing. We have also implemented the filter corresponding to a second order critically damped system,

\[
c(f_k) \mapsto {\tilde c}(f_k) = \frac{c(f_k)}{\left( 1 + i\frac{f_k}{f_c}\right)^2},
\]

which is one of the basic models for the EMG-to-force transfer function \citep{Soechting1975}. Note that the function \verb|EMG_crit2| too rectifies the EMG before filtering.

For comparison of EMG vs EMG, or (filtered) EMG vs force etc, the correlation methods are essential. Given two time series $x$ and $y$ we may define a correlation function $c_{xy}(t)$ by,

\[
c_{xy}(t) = \frac{\int_0^T {\tilde x}(u) {\tilde y}(u + t) du}{\sqrt{\int_0^T {\tilde x}(u)^2 du} \sqrt{\int_0^T {\tilde y}(u)^2 du}},
\]  

where ${\tilde x}$ is $x$ with the mean value $\bar x$ subtracted, ${\tilde x}(t) =  x(t) - \bar x$, etc. In the discrete version this is implemented by \verb|EMG_corr| employing  \verb|fft| methods. In the frequency domain a coherence function is defined by,

\[
\label{EQ:Coherency}
\text{coh}_{xy}(f) = \frac{\langle \hat{x}^\star(f) \hat{y}(f)\rangle}{\sqrt{\langle |\hat{x}(f)|^2\rangle} \sqrt{\langle |\hat{y}(f)|^2\rangle}},
\]

where $\langle \cdots \rangle$ denotes statistical averaging. This is estimated by \verb|EMG_coh| by dividing the time series into time slices of size \verb|DT| and calculating the Fourier coefficients for these slices, and finally take the averages over the blocks. \verb|EMG_corr| and \verb|EMG_coh| can be used to investigate time lags and phase shifts between signals. In the case that we have a linear relationship, $y = H \star x + n$, with a transfer function $H$ (and uncorrelated ''noise'' $n$), we would get 

\[
\mathrm{coh}_{xy}(f) = \frac{\hat H(f)}{|\hat H(f)|} = e^{i \phi(f)},
\]

where $\phi(f)$ is the phase function of the transfer function. A related time shift $\tau$ can then be obtained from $2 \pi \tau = d\phi(f)/df$.

\subsubsection{Frequency band analysis}

As is well known from musical transcription, it is convenient to study sound by how the power (intensity) is distributed over the frequency bands (pitch) as a function of time. This is useful also in basic signal analysis. The idea is to decompose signals using filter banks. Wavelets can be considered as a special realization of the idea of filter banks \citep{Vetterli1995}. An interesting hybrid method for ''intensity analysis'' of EMG has been proposed by \citep{Tscharner2000}. The idea is to divide the frequency band of interest, say one from 10 Hz to 200 Hz, into subbands centered on frequencies $f_c^{(j)}$ ($j = 1, \cdots , J$) such that the relative bandwidth $BW = \Delta f(j)/f_c^{(j)}$ scales as $1/\sqrt{f_c^{(j)}}$ over the frequency band. Here $\Delta f(j)$ is the frequency resolution of the ''mother wavelet'' $\psi$ at the center frequency $f_c^{(j)}$. In this way one can provide a distribution of signal power among the frequency bands. The power in the frequency band $j$ at time $t$ is given by $\left|c_j\right|^2$ where,

\[
	c_{j}(t) = \int \bar{\psi}_j (u - t) x(u) du,  
\] 

and ${\psi}_j$ is the wavelet centered at $f_c^{(j)}$. This differs from the recipe in \citep{Tscharner2000} but that is mainly because we use here the full complex coefficient. We present here a modification \citep{Borg2003} which is based on the Morlet function which in frequency space is given as,

\[
\hat{\psi}(f_c, \alpha, f) = \exp\left( - \frac{2 \pi^2}{\alpha f_c} (f - f_c)^2 \right).
\]

For the center frequencies we select, following von Tscharner,

\[
f_c^{(j)} = \frac{1}{s}(q+j)^2 \quad (j = 0, \cdots, J-1),
\]

determined by the parameters $s$ (''scale'') and $q$. The implementation is given by \verb|EMG_morvt| which is again based on using the \verb|fft| transformation. This function returns a \verb|list| where \verb|powc| refers to the matrix of the $c^{(j)}[i]$ coefficients, \verb|freqc| to the array with the center frequencies, and \verb|freqm| contains an estimate of the instantaneous mean frequency calculated as the average of the center frequencies weighted with the power coefficients $\left|c_j\right|^2$.

\subsubsection{Multi resolution analysis, MRA}

In the above examples we have relied on the basic libraries that belong the the default setup of the R system. In the next example we will take advantage of a library that provides functions for discrete wavelet analysis. There is for instance a package appropriately named \verb|wavelets| (by Erich Aldrich). In order to install it one enters the command

\begin{verbatim}
install.packages("wavelets")
\end{verbatim}

which will look up a depository and ask you to download the package. When successfully installed it can be loaded by the command

\begin{verbatim}
library(wavelets)
\end{verbatim}

The command \verb|library()| with empty argument will show the packages installed on your system. Information about the package \verb|wavelets| can be obtained by the command

\begin{verbatim}
 help(package = "wavelets")
\end{verbatim}
 
or \verb|??wavelets|. In multi resolution analysis (MRA) we repeatedly apply low- and high-pass filters to a discrete time series which thus can be decomposed into fine and coarse grained parts. The simplest example is the Haar filter. If $x = (x_1, x_2, \cdots)$ then Haar low pass and high pass filters produce the series,
$a = (a_1, a_2, \cdots)$ and If $b = (b_1, b_2, \cdots)$, with

\begin{align*}
&a_i =  \frac{x_{2i} + x_{2i-1}}{\sqrt{2}}, \\
&b_i =  \frac{x_{2i} - x_{2i-1}}{\sqrt{2}}. \\
\end{align*}

The averaging procedure produces a coarse grained sum version $a$, while $b$ contains the detail. Symbolically the decomposition can be written $x = (a|b)$. This procedure can be repeated taking $a$ as an input for the decomposition procedure. In this fashion we obtain

\[
x = (a^J|b^J|b^{J-1}| \cdots |b^1),
\] 

for a decomposition of order $J$. The $k$:th level detail coefficients $b^k$ represent information about the changes in the times series on a time scale proportional to $2^k$.

The function \verb|EMGx_mra()| is a wrapper for \verb|mra| in the \emph{wavelet} package. A new feature here is that the function \verb|mra| returns a \verb|class| object with \verb|slot|s whose names can be accessed by the function \verb|slotNames|. For instance the detail coefficients have the name \verb|D| and the sum coefficients the name \verb|S|. If

\begin{verbatim}
res <- mra(X)
\end{verbatim}

then the vectors with the coefficients are accessed as 

\begin{verbatim}
res@D[[j]], and res@S[[j]],
\end{verbatim}

for  the level $j$. The original data can be obtained as a sum of the decomposition,

\begin{verbatim}
X = res@D[[1]] + res@D[[2]] + ... + res@D[[J]] + res@S[[J]].
\end{verbatim}

Thus \verb|res@D[[j]]| reflects the signal content on a time scale of the order $2^j \cdot f_s^{-1}$ where $f_s$ is the sampling rate.

\subsubsection{Batch processing}

As the number of data files grow it is important to be able to process them in one row. This kind of batch processing can be simply implemented in R. We will assume that have a set of data files \verb|name1.asc|, ... , \verb|nameN.asc|. One can collect these paths of these files into \verb|filelist.asc| and write a R-script which opens each of these files for processing. One thing to remember is that the file paths must be on the Unix format using \verb|/| (or \verb|\\| ) instead of \verb|\|. The files can also be selected interactively by using the \verb|tk_choose.files()| function,

\begin{lstlisting}
library(tcltk) # load the tcltk package
Filters <- matrix(c("EMG data", ".asc", "All files", "*"),
                  2, 2, byrow = TRUE)

if(interactive()) filelist <- tk_choose.files(filter = Filters)
\end{lstlisting}

This will open the ''Select files'' dialogue and put the selected files into the \verb|filelist| variable (with file paths on the Unix format). The following snippet is a simple example which opens the files in the \verb|filelist| and plots the first column to a pdf-file, and writes the standard deviation to a text-file.

\begin{lstlisting}
# filelist -- contains paths to asci files with EMG data
outputpdf <- "C:/EMGanalysis.pdf"  # output graphs to this file
outputtxt <- "C:/EMGanalysis.txt"  # output text/numbers to this file

pdf(outputpdf) # starts the pdf driver and opens the output pdf-file
fp <- file(outputtxt, "w") # opens text-file for writing
n <- length(filelist) 

for(i in 1:n){
	EMG <- read.table(filelist[i], header = FALSE)
	title <- paste("Data from ", filelist[i])
	# this one goes to the text file -->
	cat("Standard deviation = ", sd(EMG$V1), 
	" for data EMG$V1 in file ", 
		filelist[i], "\n", file = fp) 
	# this one goes to the pdf file -->	
	plot(EMG$V1, main = title, type = "l") 
	
}

close(fp) # closes the text file
dev.off() # closes the pdf driver
\end{lstlisting}

It illustrates how one reads the data, opens a file for writing, where the writing to the file is performed using the function \verb|cat|. (It computes the standard deviation of the time series using the function \verb|sd| and writes it to the file.) This example is easily generalized to more complicated processings. 

\subsubsection{Linking C code with R}

Those who have experience with writing codes in C may find it useful to make some
functions written in C available in R. Basically one writes the function to be of type \verb|void| with the results as an argument of the function. All the arguments of the function must be pointer variables. As header files one includes \verb|R.h| and \verb|Rmath.h| (the latter is needed if one uses mathematical function from R in the C code). The source is compiled with the command

\begin{verbatim}
R CMD SHLIB my_c_code.c
\end{verbatim}

which in Windows produces a dll file \verb|my_c_code.dll| which is then loaded in R by \verb|load("my_c_code.dll")| (the details differ a bit in Unix systems). If, for instance, the c-function is of the type (computes the square)

\begin{verbatim}
void foo(int *input, int *result){
	 int x;
	 x = *input;
	*result = x*x;
}
\end{verbatim}

then one would call it in R by (with eg 5 as the input value)

\begin{verbatim}
s <- .C("foo", input = as.integer(5), result = as.integer(0))
\end{verbatim}
 
where the output \verb|s| is a list in which eg \verb|s$result| in this case contains the value computed by the function. For more details on linking C with R, see eg \citep{Ripley2008,McGlinn2011}.

\subsubsection{Data reduction -- PCA}

As an illustration of matrix methods in R we take the example of data reduction using \emph{Principal component analysis} (PCA). Suppose we have $M$ trials and each trial corresponds to a numerical data vector $\mathbf{x}^{(i)}$ of length $N$ $(i = 1, 2, \cdots , M)$. Thus the total data set corresponds to $M$ points in an $N$-dimensional space. The basic idea of PCA is to find a unit vector $\mathbf{q}$ in the $N$-dimensional data space which defines the line which lies closest to all the data points. (By subtracting mean values we can assume that the data set is centered at the origin.) 
For a mathematical definition one can look for the vectors $\mathbf{q}$ making the sum of squared projections on $\mathbf{q}$,

\begin{align*}
\sum_{i} (\mathbf{q} \cdot \mathbf{x}^{(i)})^2,
\end{align*}

extreme with the condition that $|\mathbf{q}| = 1$. This leads to an eigenvector equation of the form,

\begin{align*}
&S \mathbf{q} = \lambda \mathbf{q}, \quad \mbox{where (the division by $M$ is for convenience)} \\
&S_{jk} = \frac{\sum_{i = 1}^{M} x_j^{(i)} x_k^{(i)}}{M}.
\end{align*}

Thus we are lead to the problem of calculating eigenvectors ($\mathbf{q}$) and eigenvalues ($\lambda$) for the ''correlation matrix'' $S = D D^t/M$, where the data matrix $D$ is defined as
$D = [\mathbf{x}^{(1)}, \cdots , \mathbf{x}^{(M)}]$ and $D^t$ is the transpose of $D$. In R, if \verb|x_i| are data arrays of length $N$, the data matrix $D$ and the correlation matrix $S$ can be defined by

\begin{lstlisting}
	D <- cbind(x_1,x_2,  ..., x_M)
	S <- D %*% t(D)/M
\end{lstlisting}

Note that for the matrix multiplication we use the multiplication symbol \verb|%*%| and for the transpose of \verb|D| we use \verb|t(D)|. In R the eigenvectors and eigenvalues of a symmetric square matrix \verb|S| are obtained by the function \verb|eigen|,

\begin{lstlisting}
	out <- eigen(S, symmetric = TRUE)
\end{lstlisting}

The output \verb|out| is a list object with the eigenvectors in \verb|out$vectors| and
the corresponding eigenvalues in \verb|out$values|. The eigenvalues are ordered form largest to smallest. The projection of the data on the first eigenvectors (with largest eigenvalues) are called the \emph{principal components}. In case the first $K$ eigenvalues \verb|out$values[,1:K]| dominate the rest, we can use the projection of the data on the vectors \verb|out$vectors[,1:K]| as a substitute for the full dimensional data. That is, we use a $K$-dimensional subspace of the original data space. Examples of using PCA  in signal analysis can be found in the review \citep{Castelli2007}. 
\par

Above was an example of \emph{intra-trial} analysis were we sum over the trials. In a dual \emph{inter-trial} analysis we instead sum over data dimension and look for correlations between trials. For instance, we might have two groups of trials, with fatigued and with non-fatigued subjects, and are interested in whether these two groups can be separated by the data. One approach is again to use PCA but in this case the correlation matrix is given by \verb|U| instead of \verb|S| above,      

\begin{lstlisting}
	U <- t(D) %*% D/N
\end{lstlisting}

but otherwise the analysis follows the same pattern. For some applications to EMG see \citep{Tscharner2002}.

\subsection{Statistics}

R is by definition a statistics software package whence all the well-known, and many less well-known, statistical procedures are implemented. Important sub\-topics are descriptive statistics, statistical testing, and modeling data. Since our emphasis here is on signal processing we will not go into the statistical methods. At the very basic level we have, for instance, \verb|hist(X)| which computes and plots a histogram of numerical data \verb|X|, while \verb|plot(ecdf(X))| first calculates the empirical cumulative distribution function (ecdf) and the plots the result. The \emph{Student test} is performed by \verb|t.test| and, for instance,  \verb|t.test(X, mu = 2)| computes the $p$-value for the mean to differ from 2, and the 95\% confidence interval for the mean of \verb|X|. 

For an introduction to statistical analysis using R we recommend \cite{Everitt2010} which is provided with a R-package \verb|HSAUR2| that contains the codes and the data sets.

\subsubsection{Literate statistical analysis}

One way to facilitate the reproducibility of data analysis is to embed the R scripts directly into the report which may be written with MS Word, OpenOffice, \LaTeX, HTML, etc. This is called literate statistical analysis or programming after Donald Knuth \citep{Knuth1992}. It is a way of combining the data, the code and the documentation in one place. We may mention here two packages (both work toegether with \textsc{RStudio} and \textsc{Tinn-R}), \textsc{Knitr} and \textsc{Sweave} (developed by Yihui Xie and Friedrich Leisch respectively), which can embed R code chunks and inline commands in Markdown and {\LaTeX} documents. For instance, in Markdown language the R code chunks are separated by \verb|```{r}| and \verb|```|. Example:   

\begin{verbatim}
	This is simple text.
	Next comes an R code chunk.
	
	```{r my_first_chunk, echo = TRUE}
		rand <- rnorm(1)
	```
	
	```{r my_first_plot, echo = TRUE}
		plot(rnorm(100))
	```
	My random number is `r rand`.
	
\end{verbatim}

The inline commands are separated by \verb|`r| and \verb|`| (in the example \verb|`r rand`| prints the random number computed in the chunk by \verb|rnorm|). Markdown is a simple formatting syntax invented by John Gruber for creating web pages by converting the text to HTML, see \url{http://en.wikipedia.org/wiki/Markdown}. One point about the plotting: The image will be embedded in the HTML document, so there is no need to supply an extra image file with the HTML document. 
When embedding R in {\LaTeX} the code chunks are separated by \verb|<<>>=| and \verb|@|, and the inline command by \verb|Sexpr{| and \verb|}|. Example:

\begin{verbatim}	
	\documentclass{paper}
	\begin{document}
	
	This is simple {\LaTeX} document.
	Next comes an R code chunk.
	
	<<my_first_chunk, echo = TRUE>>=
			rand <- rnorm(1)
	@
	
	Next we show a plot.
	
	\begin{figure}
	<<fig=TRUE, echo=FALSE>>=
		plot(rnorm(100))
	@
	\caption{My plot}
	\end{figure}
	
	And my random number is \Sexpr{rand}.
	
	\end{document}	
\end{verbatim}

From this \textsc{Sweave} in R generates the {\LaTeX} source that can be compiled in the normal way. In \textsc{RStudio} one can compile directly to the pdf format. For informative lectures on using \textsc{Knitr} and \textsc{Sweave} see \cite{Peng2013,Ziegenhagen2011}. 

\section{Conclusions}
We have given a brief introduction to some basic features of the R software used as a tool for analyzing and displaying EMG data, and biosignal data in general. Using R it is easy to document the exact procedures employed in analyzing the data so that it can be replicated by other researchers. A next level would be to develop a dedicated \textsc{Remg} package with tools covering various aspects of EMG and related kinesiological data and signals (MMG, ECG, etc). Such a package could be supplied with a representative set of data for testing and demonstrating the analysis methods. Finally we would also like to emphasize the usefulness of R in teaching basic data processing and visualization methods to biomechanics students.

\section*{Acknowledgments}
The author thanks Maria Finell for gathering the EMG- and balance data used in the examples (supplementary material). He is also indebted to W Jeffrey Armstrong for exchanges on the ''intensity analysis'' which has resulted in an update of the \verb|EMGfuns.R| file.

\section*{Supplementary material}

The file \verb|data.bal| contains quiet standing balance COP-data in ASCI format. First column contains COP X, second column contains COP Y, both in millimeters. The third column contains total vertical force (Newton). The columns are \verb|tab|-separated (\verb|\t|). Sampling rate is 100 Hz. The file \verb|data.emg| contains the EMG-data in ASCI format. The columns contain data from the muscles Tibialis anterior (right), Lateral Gastrocnemius (right), Medial Gastrocnemious (right), Tibialis anterior (left), Lateral Gastrocnemius (left), Medial Gastrocnemious (left), all sampled at 1000 Hz. The file \verb|emg_analysis.R| is the R-script which demonstrates a few basic analyzing methods with the EMG and the balance data. The script either produces a \verb|pdf| report (\verb|result.pdf|), or shows the results on the console, depending on the setting of the variable \verb|report| (\verb|TRUE| or \verb|FALSE|). The file \verb|EMGfiles.R| contains a script which demonstrates how to set up batch processing. The file \verb|EMGfuns.R| contains the basic R-scripts (functions) for analyzing EMG which are employed by \verb|emg_analysis.R|. The files are by default assumed to reside in the archive \verb|C:/EMGR|. The files can be downloaded as \url{http://luna.chydenius.fi/~frborg/emg/EMGR.zip}, and also from the arXiv site as supplementary material.

\bibliography{emg_analysis}
\bibliographystyle{spbasic}


\end{document}